\documentclass[useAMS,usenatbib]{mn2e}
\usepackage {graphicx}
\usepackage{natbib}
\usepackage{aas_journals}
\usepackage{color}
\usepackage{amsmath}

\title[WISE-2MASS infrared galaxies]{Star-galaxy separation strategies for WISE-2MASS all-sky infrared galaxy catalogs}
\author[A. Kov\'acs and I. Szapudi]{Andr\'as Kov\'acs$^{1,2}$, Istv\'an Szapudi$^{3}$\\
$^1$ Institut de F\'isica d'Altes Energies, Universitat Aut\'onoma de Barcelona, E-08193 Bellaterra (Barcelona), Spain\\
$^2$ MTA-ELTE EIRSA "Lend\"ulet" Astrophysics Research Group, 1117 P\'azm\'any P\'eter s\'et\'any 1/A Budapest, Hungary\\
$^3$ Institute for Astronomy, University of Hawaii 2680 Woodlawn Drive, Honolulu, HI, 96822, USA}

\begin{document}

\date{Submitted 2014}

\pagerange{\pageref{firstpage}--\pageref{lastpage}} \pubyear{2012}

\maketitle

\label{firstpage}
\begin{abstract}
We combine photometric information of the WISE and 2MASS all-sky infrared databases, and demonstrate how to produce clean and complete galaxy catalogs for future analyses. Adding 2MASS colors to WISE photometry improves star-galaxy separation efficiency substantially at the expense of loosing a small fraction of the galaxies. We find that $93\%$ of the WISE objects within $W1<15.2$ mag have a 2MASS match, and that a class of supervised machine learning algorithms, Support Vector Machines (SVM), are efficient classifiers of objects in our multicolor data set. We constructed a training set from the SDSS PhotoObj table with known star-galaxy separation, and determined redshift distribution of our sample from the GAMA spectroscopic survey. Varying the combination of photometric parameters input into our algorithm we show that $W1_{\rm WISE} - J_{\rm 2MASS}$ is a simple and effective star-galaxy separator, capable of producing results comparable to the multi-dimensional SVM classification. We present a detailed description of our star-galaxy separation methods, and characterize the robustness of our tools in terms of contamination, completeness, and accuracy. We explore systematics of the full sky WISE-2MASS galaxy map, such as contamination from Moon glow. We show that the homogeneity of the full sky galaxy map is improved by an additional $J_{\rm 2MASS}<16.5$ mag flux limit. The all-sky galaxy catalog we present in this paper covers $21,200$ square degrees with dusty regions masked out, and has an estimated stellar contamination of $1.2\%$ and completeness of $70.1\%$ among 2.4 million galaxies with $z_{med}\approx 0.14$. WISE-2MASS galaxy maps with well controlled stellar contamination will be useful for spatial statistical analyses, including cross correlations with other cosmological random fields, such as the Cosmic Microwave Background. The same techniques also yield a statistically controlled sample of stars as well.
\end{abstract}


\section{Introduction}

In recent years, sky surveys have been producing astronomical data with a rapidly accelerating
pace resulting in what is commonly called the "data avalanche''. The large quantity of data necessitates automated algorithms for filtering, photometric selection, and estimation observables such as redshifts. Each object in a catalog has multiple properties, thus algorithms have to explore high-dimensional configuration spaces in large, often connected, databases. Such high dimensional spaces can be effectively explored with machine learning techniques, such as the support vector machines (SVM) used in our work.

When analyzing an object catalog, the most fundamental, and often most challenging task is star-galaxy (possibly QSO) separation. A simple separator between stars and galaxies is a morphological measurement, where extended sources are classified as galaxies \citep{vasco2011}. Morphology, however, looses its power at fainter magnitudes, a problem for wide-field surveys, e.g.,  Pan-STARRS \citep{kaiser2010}, Euclid \citep{euclid}, BigBOSS \citep{bigboss}, DES \citep{DES}, and LSST \citep{LSST}. 
At the fainter end, the most widely used tools for object classification are color-color diagrams: different types of objects will appear in different regions  according to the shape of their spectral energy distribution. Classification methods based on infrared color-color selection were employed for star-galaxy separation \citep{Pollo2010} or for finding special classes of sources, such as high/low redshift QSO, AGN, starburst galaxies, or variable stars \citep[e.g.,][and references therein]{richards2002,chiu2005,stern2012, brightman2012}.

Applications of SVMs are widely used for data mining and analysis. SVMs are relatively easy to implement, simple to run, and they are very many-sided, thus several astronomical problems set out to use such methods to classify objects, or to perform regression analysis.
Among others, \cite{wozniak2004} analyzed variable sources with an SVM in a 5 dimensional parameter space including period, amplitude and three colors, \cite{huertas2009} analyzed morphological properties of infrared galaxies using SVM with 12 parameters, while \cite{solarz2012} created a star-galaxy separation algorithm based on mid and near-infrared colors. Recently, \cite{malek2013} used VIPERS and VVDS surveys to perform object classification into 3 groups: stars, galaxies, and AGNs. A similar study by \cite{saglia2012} also used an SVM as 3-type-classifier. Their method was developed for the Photometric Classification Server for the prototype of the Panoramic Survey Telescope and Rapid Response System 1 \citep[Pan-STARRS1]{kaiser2010},.

While star-galaxy separation can be performed from WISE colors alone \citep{tomo,KovacsEtAl2013} it is at the expense of  severe cuts that still are sensitive to contamination to from the Moon and necessitate complex masks. As we show later, adding 2MASS observations removes artificial features from the WISE data.
With several open source implementations and computationally modest cost \citep{fadely2012}, we set out to use the SVM algorithm for separating stars and galaxies in the matched WISE-2MASS photometric data. Our principal goal is to create a clean catalog of galaxies observed by WISE and 2MASS suitable for large scale structure and cross-correlation studies. At the same time, we will show that our selection algorithms  are suitable for producing clean stellar samples as well. The galaxy maps we create are useful for cross-correlation studies, such as Integrated Sachs-Wolfe (ISW) measurements, and galaxy-Cosmic Microwave Background (CMB) lensing correlations, while the large data sets of stars may constrain stellar streams and Galactic structure in general. 

The paper is organized as follows. Datasets and algorithms are described in Section 2, while our results are presented in Section 3, with detailed discussion, comparisons, and interpretation. 

\section{Datasets and methodology}

We combine measurements of two all-sky surveys in the infrared, Wide-Field Infrared Survey Explorer \citep[WISE]{wise} and the Point Source Catalog of the 2-Micron All-Sky Survey \citep[2MASS-PSC]{2mass}. We use photometric measurements of the WISE satellite, which surveyed the sky at four different wavelengths: 3.4, 4.6, 12 and 22 $\mu$m (W1-W4 bands). Following \cite{tomo} and \cite{KovacsEtAl2013} we select sources to a flux limit of W1 $\leq $15.2 mag to have a fairly uniform dataset.

We then add 2MASS J, H and $K_{s}$ magnitudes conveniently available in the WISE catalog, where $93\%$ of the WISE objects with W1 $\leq $ 15.2 mag have 2MASS observations. We find, however, lower matching rates for a WISE-2MASS sample at fainter W1 cuts.

Note that the WISE W1 magnitude limit we define is lower than the $5\sigma$ detection limit for W1. However, this selection cut makes comparisons to previous WISE catalogs easier, and helps to avoid large-scale inhomogeneities (caused by moon glow effects) which potentially contaminate deeper WISE galaxy catalogs \citep{KovacsEtAl2013}.

We note that these choices allow us to produce a catalog deeper than the 2MASS Extended Source Catalog \citep[2MASS]{jarrett2000}, as proper identification of fainter 2MASS objects becomes possible.

To apply machine learning techniques, one needs to identify a "training set'', a set of objects with known classification.  We choose a smaller region of Stripe 82 in the Sloan Digital Sky Survey Data Release 7 \citep[SDSS-DR7]{sdss}, deeper than our catalog and located at $327.5 < RA < 338.5$ and $-1.25 < Dec < 1.25$. We performed the cross-matching with the KD-Tree \citep{kdtree} algorithm as implemented in the {\tt python} package {\tt scipy}.  We found an SDSS match for 99.4$\%$ for the 46,749 WISE-2MASS objects using a 1" matching radius. We have found multiple counterparts for $0.03\%$ among 46,463 objects, and chose the nearest neighbor as a real counterpart. We estimated the accidental matching rate by generating random positions of WISE-2MASS source density, finding $0.1\%$. This suggests no meaningful effect on the training sample.
As a further refinement, we applied a W1 $\geq $ 12.0 magnitude cut to avoid potentially problematic SDSS classification, and to mark out the galaxy locus in color-magnitude space. See Fig.~\ref{brights} for ratification.

As an exploratory test, we downloaded 2MASS XSC data \citep{jarrett2000} from the same coverage, finding 1,195 galaxies. A deeper WISE-2MASS catalog without extra flux cut in $J$ band contains 5,922 objects classified as a galaxy in SDSS PhotoObj table. We will show that the fraction of the properly identified galaxies reaches $\sim 78.6 \%$ with $\sim 1.8 \%$ star contamination (see Table \ref{t1}), even with our simplest algorithms. We thus are able to broaden 2MASS XSC significantly.

The redshift distribution of the matched WISE-2MASS-SDSS objects classified as galaxies is provided by matching with the Galaxy and Mass Assembly \citep[GAMA]{gama} spectroscopic dataset, at the full GAMA coverage of 144 $deg^{2}$. We preformed a nearest neighbor search using $1"$ as a matching radius, and found a pair for $97\%$ of the WISE-2MASS-SDSS galaxies in GAMA Data Release 2. We have found multiple counterparts for $0.15\%$ among 8,493 objects, and chose the nearest neighbor as a real counterpart. The un-matched $3\%$ might consist of predominantly massive early-type galaxies at $z > 1$ \citep{yanetal2012}. Another possibility is that the remaining $3\%$ is populated by objects with bad SDSS classification, that actually indicates the purity of our training sample.

Next we will use the resulting multicolor catalog for object classification. Note that the training set we use is a realistic sample of the multicolor WISE-2MASS data base, since we applied the same flux cuts for all subsamples in the analysis.

\begin{figure*}
\begin{center}
\includegraphics[width=195mm]{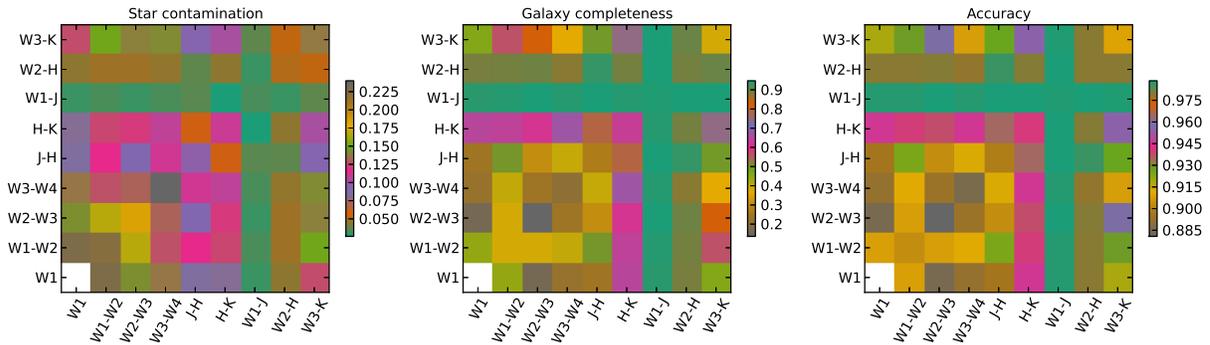}
\caption{Measures of SVM performance are presented in the case of pairwise and single usage. Color-coded maps illustrate contamination, completeness, and accuracy for every combinations. All subfigures suggest that W1-J is a dominant potency in star-galaxy separation. We note, however, that SVM failed to produce precise results using W1 alone. Every object was classified as a star with that choice, thus we excluded the W1-only case from the analysis. Combinations of W1 and other parameters, however, are preserved, as they produce valuable results.}
\label{ims}
\end{center}
\end{figure*}

\subsection{Support Vector Machines}

SVM designates a subclass of supervised learning algorithms for classification in a multidimensional parameter space. These methods include extensions to nonlinear models of the generic (linear) algorithm developed by \cite{vapnik}. SVMs carry out object classification and/or regression by calculating decision hyperplanes between sets of points having different class memberships.
A central concept of SVM learning is the training set, a special set of objects that supplies the machine with classified examples. Based on its properties, the classifier is tuned, and the hyperspace between different classes is determined. A training set of a few thousands of objects is usually suitable for simpler classification problems (see e.g. \cite{solarz2012}).

The algorithm includes a non-linear kernel function, which is used to find a hyperplane with maximum distance from the boundary to the closest points belonging to the separate classes of objects \citep{vapnik}. The kernel is a symmetric function that maps data from the input space X to the feature space F. For our analysis we chose a Gaussian radial basis kernel (RBK) function, defined as $k(x_i, x_j) = e^{ -\gamma||x_i-x_j ||2} $,where $||x_i-x_j ||$ is the Euclidean distance between $x_i$, and $x_j$. The product of the kernel function is a non-linear representation of each parameter from the input to the feature space. The RBK kernel is often used as SVM kernel function to make the non-linear feature map. We decided to use it because of its effectiveness and simplicity.

SVM offers a whole set of parametrization choices. We chose 'C-classification' because of its good performance and only two free parameters. C is the cost function, i.e. a trade-off parameter that sets the width of the margin separating classes of objects. A small margin of separation can be set with larger C, but high C values often lead to over-fitting.  Reduced C values, however, smooth the hyperplane, which can be a source of mis-classifications \citep{malek2013}.
The second parameter, $\gamma$, determines the topology of the decision surface. A low value of $\gamma$ sets a rigid, and structured decision boundary, while high $\gamma$ values indicate a very smooth decision surface with many mis-classifications.

\begin{figure}
\begin{center}
\includegraphics[width=85mm]{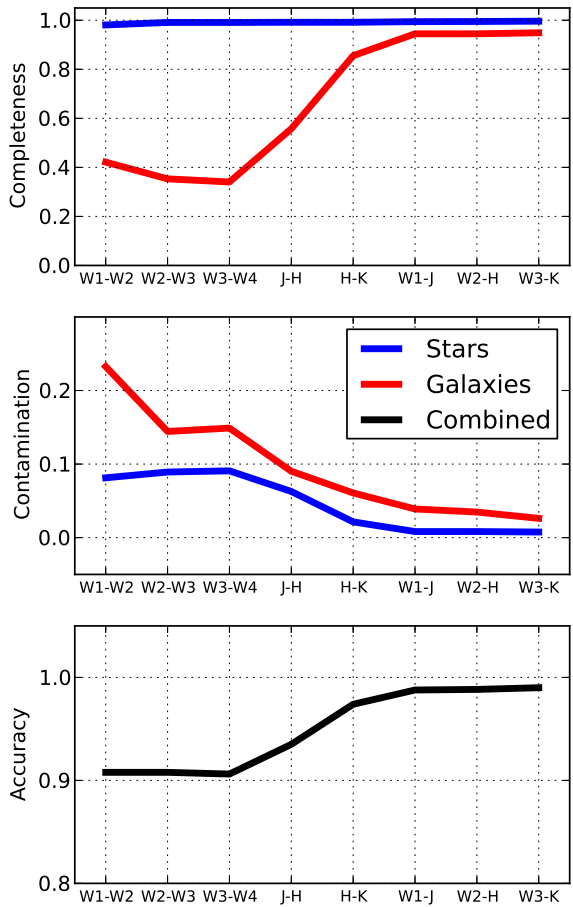}
\caption{Measures of SVM performance are shown as a function of SVM parameters. We observed upgrading trends in contamination, completeness, and accuracy for both stars and galaxies. High completeness values for the star sample can be explained by the fact that the sample is dominated by stars, thus False galaxies cannot affect star completeness significantly.}
\label{conts}
\end{center}
\end{figure}
\section{Discussion}

\subsection{SVM outputs}

We used a free software environment for SVM in {\tt python} package {\tt scikit-learn}. First we performed tests to tune both the C and $\gamma$ parameters, and found the lowest classification errors with C=10.0, and $\gamma = 0.1$. Then we proceeded to determine the optimal number of parameters for the optimal classification efficiency experimentally. We used 8,000 objects as a "training set'', and 2,000 objects for control, i.e. testing the efficiency of our algorithms. The training sample contains $13\%$ galaxy data, in order to preserve the star-galaxy ratio that we have originally found in our SDSS cross-matching.
We evoke the terminology of machine learning, and use ''True'' ($T$) and ''False'' ($F$) labels to distinguish between objects that are classified correctly, and the ones have false identification.

We also define five measures of SVM performance:
\begin{itemize}
\item Star Contamination = $\frac{F_{S}}{T_{S}+F_{S}} $
\item Galaxy Contamination = $\frac{F_{G}}{T_{G}+F_{G}} $
\item Star Completeness = $\frac{T_{S}}{T_{S}+F_{G}} $
\item Galaxy Completeness = $\frac{T_{G}}{T_{G}+F_{S}} $
\item Accuracy = $\frac{T_G+T_S}{T_G+F_G+T_S+F_S}$
\end{itemize}

We used the following set of colors/magnitudes as input parameters: $W1,W1-W2, W2-W3, W3-W4, J-H, H-K_{s}, W1-J, W2-H,$ and $W3-K_{s}$. 
Initially, we supplied SVM with all possible pairs of this set, and obtained contamination, completeness, and accuracy. The parameter $W1-J$ is an astoundingly good star-galaxy separator, as shown in Figure \ref{ims}. Either alone or combined with any other parameter, $W1-J$ guarantees the lowest stellar contamination, the highest galaxy completeness, and the highest accuracy. For instance, the stellar contamination for the combination of $W1$ and $W1-J,$ or $ H-K_{s}$ and $W1-J$ is as low as $3.1\%$, while the galaxy completeness is $93.6\%$ (see Table 1).

\begin{figure}
\begin{center}
\includegraphics[width=85mm]{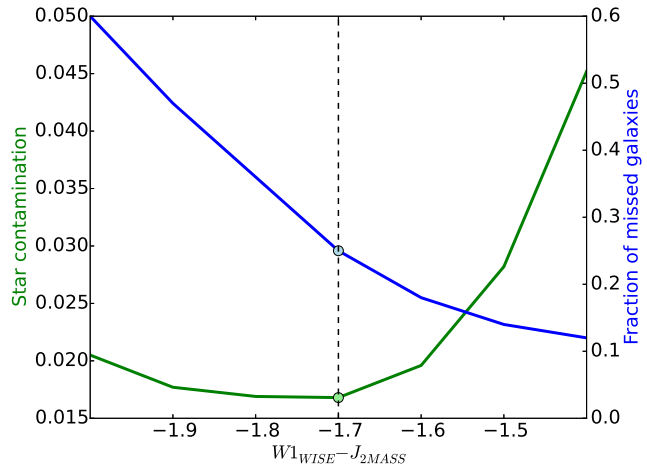}
\caption{Star contamination (blue curve, values on the left axis) and the fraction of the lost galaxies (green curve, values on the right axis)  is shown. We applied a W1-J $\leq $ -1.7 cut, and now show the consequences for the parameters of the resulting catalogs.}
\label{conts2}
\end{center}
\end{figure}

Next we supplied the SVM with more parameters. We started with $W1-W2$ alone, then added one more parameter in each step. Our findings are summarized in Figure \ref{conts}. We qualitatively confirmed our former results, namely that the combination of WISE and 2MASS parameters increases the SVM performance. For WISE colors only, galaxy completeness is at the level of $\approx 40\%$, while with all parameters it reaches $\approx 93 \%$. At the same time, stellar contamination decreased from $\approx 15\%$ to $\approx 3\%$. Finally, similar trends are seen for the accuracy parameter, that incremented by $\approx 10\%$ by adding 2MASS parameters.

As a test of possible impurities in the training sample, we randomly flipped the classification of $1\%$, $3\%$, $5\%$, and $10\%$ of the training objects, and repeated our SVM analysis with the artificially contaminated training sets. We analyzed the case of the combination of $W1-J$ and $W1$, finding $3.7\%$, $4.8\%$, $7.7\%$, and $11.2\%$ for the star contamination, respectively, while the original value with the unchanged training set was $3.1\%$.

\subsection{SVM vs. color-color and color-magnitude cuts}
Findings of the previous subsection suggest that separation of stars and galaxies can be achieved a simple cut on the W1--W1-J color-magnitude plane. Stellar contamination and galaxy completeness are then comparable to that of the multicolor SVM algorithm, but with a faster method.
Figure~\ref{conts2} shows the estimated stellar contamination and the ratio of properly identified and lost galaxies in the case of different $W1-J$ cuts. We choose $W1-J$ $\leq $ -1.7 for our purposes, as it guarantees the lowest stellar contamination, while $78.6\%$ of the galaxies can be classified as galaxies.

\begin{figure*}
\begin{center}
\includegraphics[width=155mm]{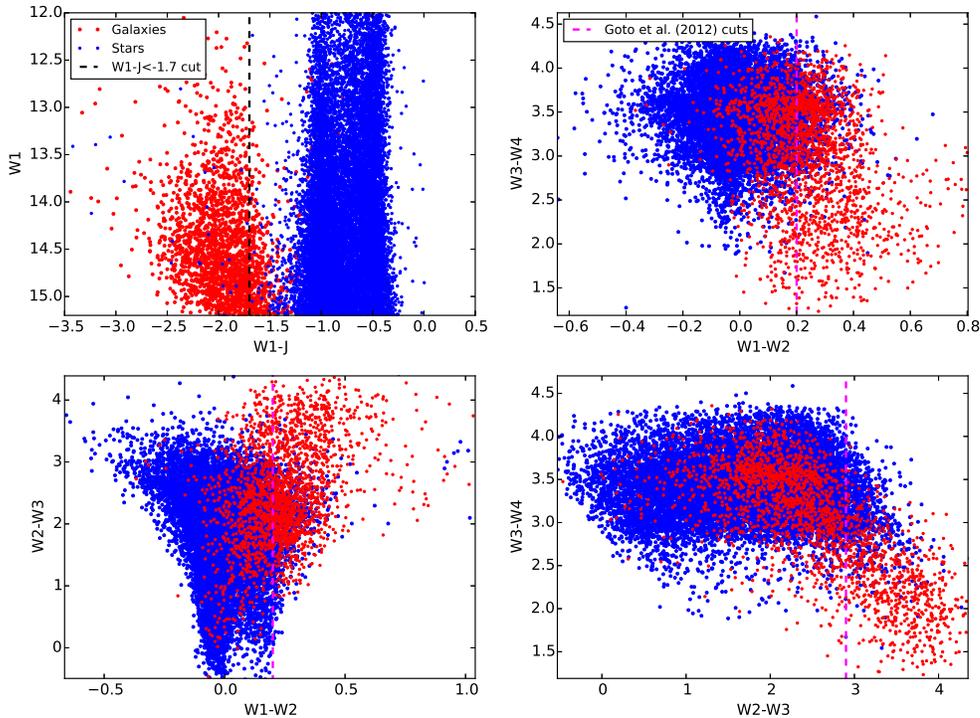}
\caption{Top left: A simple star-galaxy separator which uses only W1, and W1-J color. The separation of stars and galaxies is remarkably strong in this parameter space. Other subfigures show color-color plots of the four WISE bands. We show the special galaxy separator cuts applied by Goto et. al (2012). This result illustrates that star-galaxy separation on traditional color-color planes with linear cuts is challenging, if one wants to use a large fraction of the achievable galaxy sample.}
\label{hists}
\end{center}
\end{figure*}

Visualizing this parameter choice, we show a W1--W1-J, and WISE color-color diagrams for WISE-2MASS objects from subsample of 20,000 objects (with $13\%$ galaxy content) in Figure \ref{hists}. Classes are indicated by SDSS in this comparison. We note that a remarkable separation of stars and galaxies can be seen in the upper left of Figure \ref{hists}. 

\begin{figure}
\begin{center}
\includegraphics[width=65mm]{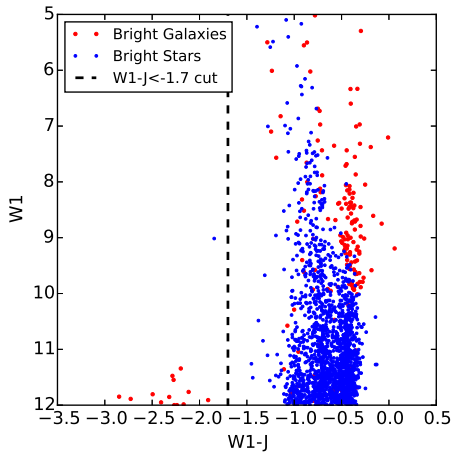}
\caption{The bright "galaxies" of W1 $\geq $ 12.0 mag with potentially bad SDSS classification are located in the stellar locus. We excluded all britht objects from the analysis, since their presence can alter the training efficiency. See text for details.}
\label{brights}
\end{center}
\end{figure}

However, the patterns we found in Figure \ref{brights} enforce a $W1$ $\geq $ 12.0 magnitude cut, as a larger subsample of SDSS "galaxies" imbedded in the definite stellar locus of this plot. This fact suggests that these objects might have been mis-classified by SDSS, and their usage is unsafe in a training set. We investigated the actual SDSS image of these "galaxy" objects, and found that they are indeed really bright stars with bad classification.
We emphasize, that neither our SVM methods nor the W1--W1-J based simple galaxy selection are not affected, since we removed these brightest objects from our sample.

\begin{figure}
\begin{center}
\includegraphics[width=70mm]{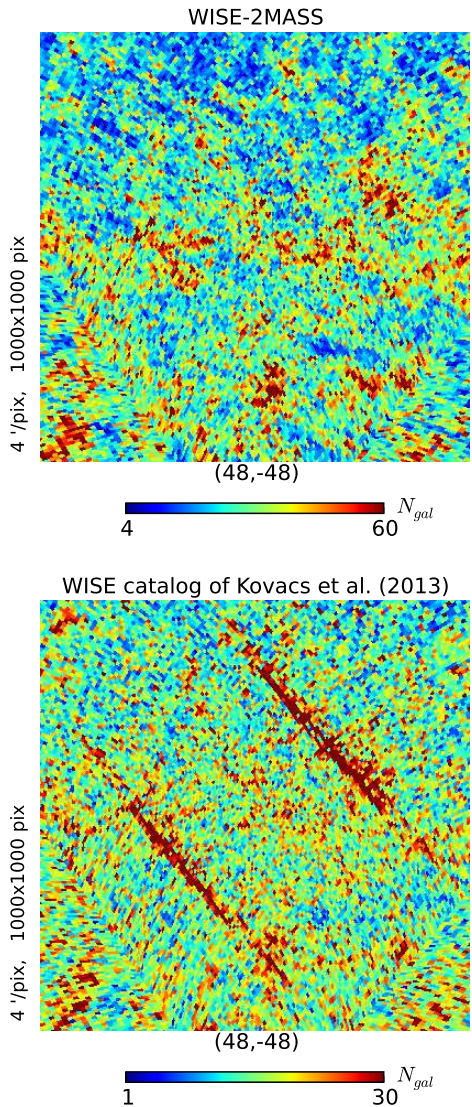}
\caption{Gnomonic projection of galaxy number counts in HEALPIX pixels at $N_{side}=128$ is shown for the sample of \citep{KovacsEtAl2013} (Top) and present approach (Bottom). Figures are centered on $\ell,b=48.0,-48.0$. }
\label{stripes}
\end{center}
\end{figure}

\begin{figure}
\begin{center}
\includegraphics[width=80mm]{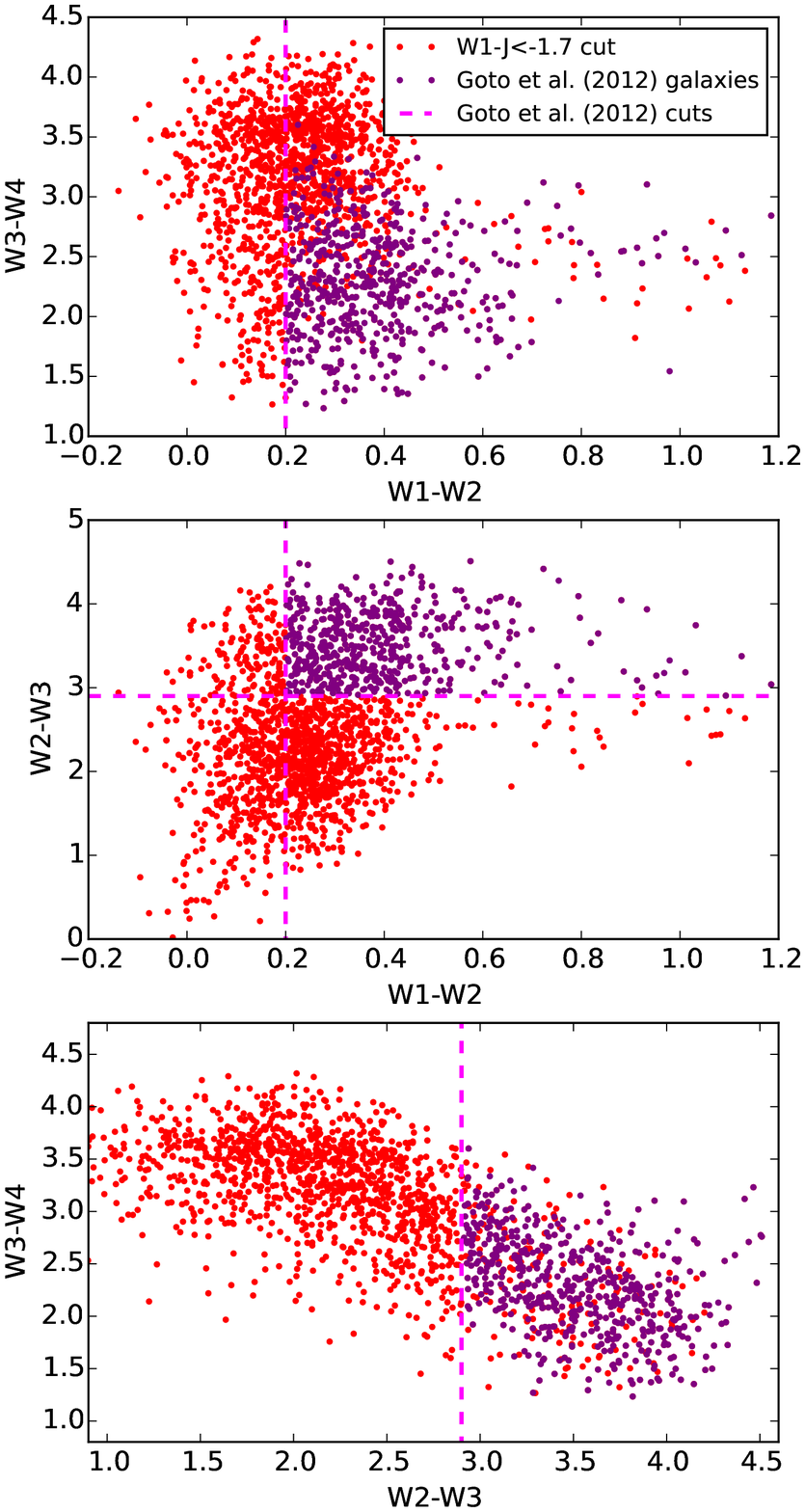}
\caption{Distributions of galaxies on color-color plane is shown for \citep{tomo} cuts and galaxies selected in present work. Significant amount of galaxies is identified properly with the W1-J $\leq $ -1.7 galaxy cut. We used $1,785$ galaxies for making this figure, i.e. all galaxies that we identified properly using our simple CC tool (see Figure \ref{hists} and Table 1 for details). This represents a $78.6\%$ galaxy completeness. }
\label{AT}
\end{center}
\end{figure}

\subsection{Further comparisons}

Next we compare our galaxy sample to that of \cite{tomo}, and \cite{KovacsEtAl2013}. While these works used all four observations of WISE, we only need W1, i.e. the one with the best quality observations. The stellar contamination we estimate for \cite{tomo}, and \cite{KovacsEtAl2013} using the SDSS classification is $7\%$, while only $1.8\%$ for our present sample applying the $W1-J$ $\leq $ -1.7 cut.
At the same time, while previous WISE-only methods produced $21\%$ galaxy completeness, presently we achieved a $78.6\%$ with the new galaxy selection criteria. SVM results reach $\approx 93\%$ completeness for galaxies, with similar stellar contamination as the $W1-J$ cut. Figure \ref{AT} summarizes our findings.

We note that the stellar contamination may be higher where the number density of stars is above the average, e.g. close to the Galactic plane, or at the Small and Large Magellanic Clouds. Among others, these regions should be masked out in order to avoid mis-classification problems.

There are other object separation algorithms in the literature, but either they are optimized for QSO-AGN selection \citep{yanetal2012}, or limited to bright magnitude cuts \citep{jarrett2011}. We argue, therefore, that a direct and detailed comparison is only possible with the results of \cite{tomo}, and \cite{KovacsEtAl2013}.

\subsection{All-sky galaxy catalog}

$W1-J$ cut appears to be a powerful tool for separating stars and galaxies, providing a fast and simple option to create a full-sky WISE-2MASS galaxy map. The simple cut can be realized by a query into the WISE-2MASS database. As W1-J $\leq $ -1.7 gives the lowest contamination according to our tests, we selected galaxies with the following query:
\begin{center}
\begin{verbatim}
w1mpro between 12.0 and 15.2 and 
n_2mass > 0 and 
w1mpro - j_m_2mass < -1.7 and 
glat not between -10 and 10
\end{verbatim}
\end{center}
where "w1mpro" is the $W1$ brightness, "n 2mass" is the number of the associated 2MASS sources, "j m 2mass" corresponds to the brightness in the $J$ band, and "glat" is the Galactic latitude coordinate.

We downloaded $\sim5$ million WISE-2MASS objects from the IRSA website\footnote{\texttt{http://irsa.ipac.caltech.edu/}}. The dataset contained $W1,W2,W3$ and $W4$ for WISE, and $J, H$ and $K_s$ for 2MASS as photometric parameters, and we also downloaded 'cc flag' values to uphold the possibility of further restrictions.

Next, we test for possible biases and systematic problems that may affect our selections. We have demonstrated that our SVM algorithms and $W1-J$ color cuts are capable of separating stars and galaxies, but the accuracy of the galaxy selection does not guarantee for instance spatial homogeneity. We have shown in \cite{KovacsEtAl2013}, that observational strategies of surveys may be harmful for the galaxy samples we wish to create, and inhomogeneities might show up as consequences of varying sensitivity or other observational effects.
\begin{figure}
\begin{center}
\includegraphics[width=85mm]{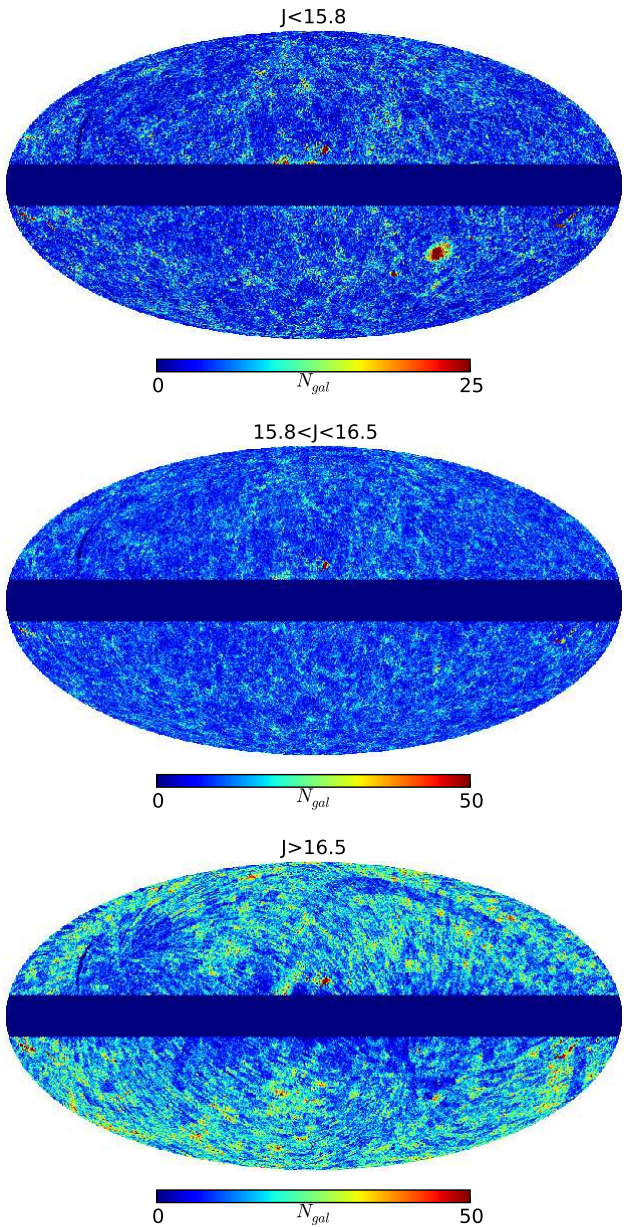}
\caption{Slices of 2MASS $J$ magnitude as probes of spatial homogeneity. We use HEALPix at $N_{side}$=128 for making these plots.}
\label{pluscut}
\end{center}
\end{figure}
First we investigate the stripe-shaped over-densities at several locations across the sky we found in \cite{KovacsEtAl2013}. These artifacts caused by moon-glow are significantly reduced in the new catalog, as shown on the upper panel of Figure \ref{stripes}. As it was pointed out by \cite{KovacsEtAl2013}, stripes are associated with the scanning strategy of the WISE survey. Different WISE bands have different sensitivities and sky coverage, therefore affect the uniformity of a full sky sample through moon-glow contamination. \cite{KovacsEtAl2013} handles this issue with special moon-contamination mask using the 'moonlev' flag of the WISE database. The stripes are not present in a $W1-J$ $\leq $ -1.7 selected dataset.

The next probe is the test of homogeneity across the sky, in particular the possible gradient in the density field as a function of Galactic latitude, as reported in \cite{KovacsEtAl2013}. We assume that the limit of $W1<15.2$ for WISE is conservative enough, although we do not apply any limitations to 2MASS data. Naturally, there is a correlation between the magnitudes of objects observed in WISE and 2MASS, but the cut of $W1-J<-1.7$ does not automatically guarantee high signal-to-noise ratio for 2MASS as well.
Therefore we divide our sample into three slices of 2MASS $J$ brightness. We use $J<15.8$ as a first limit that is the corresponding $10\sigma$ photometric detection limit for 2MASS PSC objects. This would result a dramatic loss of objects, as only $\sim 800,000$ galaxies remain in the sample. As our goal is to broaden the relatively shallow 2MASS XSC galaxy sample, we empirically obtain a higher magnitude limit for $J$ with experiments. We experimentally found that a $J<16.5$ cut effectively removes a significant amount of inhomogeneous data from our catalog. The $J>16.5$ map, however, is strongly affected by spatial variations of the sensitivity of the 2MASS PSC data, as shown in Figure \ref{pluscut}. We thus remove these objects from the analysis.

\begin{figure}
\begin{center}
\includegraphics[width=90mm]{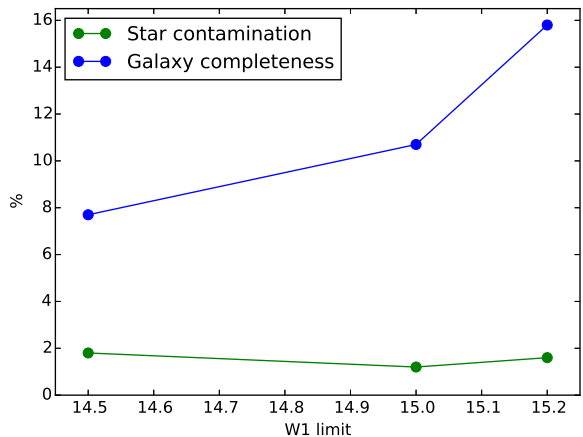}
\caption{Star contamination and galaxy completeness differences for SVM and CC methods as function of W1 magnitude limit (we use the case of an extra $J<16.5$ mag cut for W1=15.2). SVM methods result in higher contamination, but increased completeness in all cases.}
\label{CC_SVM}
\end{center}
\end{figure}

We test the SVM performance and color cuts applying different flux limits for $W1$ and $J$. In SVM analysis, we test for the case of the $W1$ and $W1-J$ parameter pair in order to use the same information as used in the case of color cuts. Our findings are summarized in Table~\ref{t1}, and in Figure \ref{CC_SVM}.

\begin{figure*}
\begin{center}
\includegraphics[width=140mm]{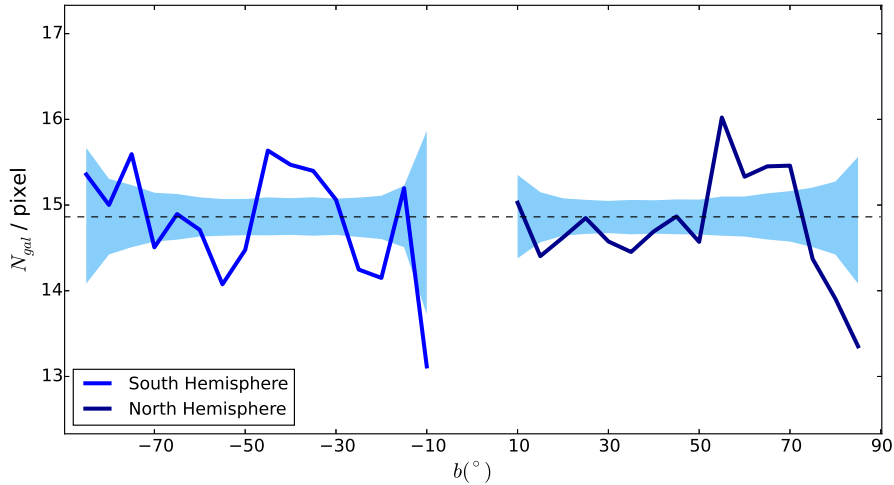}
\caption{Galaxy number count measurements in rings as a function of Galactic latitude $b$ for the $J<16.5$ galaxy map. Horizontal dashes indicate the average number of galaxies per pixel in the full map, outside the mask we constructed. Light blue shaded regions indicate Poisson errors for the measurement. We attribute the largest fluctuations to the presence of the largest superstructures in the local Universe. Note that these fluctuations are at the $\sim 2\sigma$ level, except the one at $b\approx 60^{\circ}$. Possibly these fluctuations can be lowered using a more sophisticated mask which effectively removes further potentially contaminated regions.}
\label{gradi}
\end{center}
\end{figure*}

\begin{table}
\centering
\caption{Star contamination and galaxy completeness as a function of flux limits and analysis methods (CC="$W1-J<-1.7$ color cut"). We apply a $W1>12.0$ lower flux cut in every cases. See text for details. The last column shows the expected galaxy number counts assuming a mask which leaves 21,200 ${\rm deg}^{2}$ unmasked.}
\begin{tabular}{@{}lcclll} \hline Method &  W1 & J & $\frac{F_{S}}{T_{S}+F_{S}} $ & $\frac{T_{G}}{T_{G}+F_{S}} $& $N_{gal}$ \\
\hline SVM & $<14.5$  & - & $3.4\%$ & $92.1\%$ &$1.2\cdot 10^{6}$ \\
\hline CC & $<14.5$  & - & $1.6\%$ & $84.4\%$ & $1.1\cdot 10^{6}$ \\
\hline SVM & $<15.0$  & - & $2.6\%$ & $93.4\%$ &  $2.3\cdot 10^{6}$\\
\hline CC & $<15.0$  & - & $1.4\%$ & $82.7\%$  & $2.1\cdot 10^{6}$\\
\hline SVM & $<15.2$  & - & $3.1\%$ & $93.6\%$  & $6\cdot 10^{6}$\\
\hline CC & $<15.2$  & - & $1.8\%$ & $78.6\%$ & $5\cdot 10^{6}$ \\
\hline SVM & $<15.2$  & $<16.5$ & $2.8\%$ & $85.9\%$ & $3\cdot 10^{6}$\\
\hline CC & $<15.2$  & $<16.5$ & $ 1.2\%$ & $ 70.1\%$ & $2.4\cdot 10^{6}$\\
\hline
\end{tabular}
\label{t1}
\end{table}

We note that we observe many apparent over densities near the Galactic plane, and Large and Small Magellanic Clouds are visible in the lower panel of Figure \ref{pluscut}. This finding reflects the fact that our star-galaxy separation tools fail more frequently at regions of high stellar density, as estimated and expected. We argue, however, that both the $J<15.8$ map, the $15.8<J<16.5$ map, and their union effectively trace large scale structure, and do not contain significant large scale inhomogeneities. The resulting galaxy map does not suffer from strange stripe-shaped over-densities, and confirms our finding presented in Figure \ref{stripes}. We note that this additional brightness cut shifts the median redshift of the sample from $z \approx 0.17$ to $z \approx 0.14$.

We further probe the uniformity our catalog by performing tests on possible gradients in galaxy number counts as a function of Galactic latitude, finding no such effect in the WISE-2MASS map. Figure \ref{gradi} illustrates our findings.

\begin{figure}
\begin{center}
\includegraphics[width=85mm]{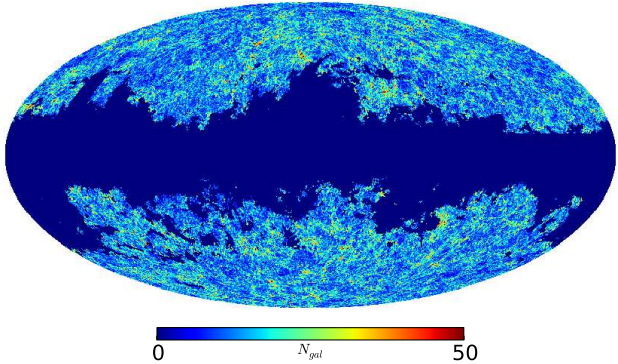}
\caption{Our example WISE-2MASS galaxy number count map applying the extra $J<16.5$ mag cut, and the corresponding mask that we constructed. We use HEALPix at $N_{side}$=128.}
\label{density}
\end{center}
\end{figure}

Finally, we construct a mask to exclude potentially contaminated regions near the Galactic plane using the dust emission map of \cite{SchlegelEtal1998}. We mask out all pixels with $E(B-V) \geq 0.1$, and regions at galactic latitudes $|b| < 10^{\circ}$, leaving 21,200 ${\rm deg}^{2}$ for our purposes. Unmasked regions in the galaxy map are corrected for Galactic extinction using the same dust map provided by \cite{SchlegelEtal1998}. We use $A_{\rm WISE}/E(B-V)=0.18$ and $A_{\rm 2MASS}/E(B-V)=0.72$ coefficients estimated by \cite{Yuan2013}.
The final all sky galaxy map is shown in Figure \ref{density}.

We note that the star-galaxy separation methods we developed are useful for selecting stellar samples as well. For instance, a $W1-J$ $\geq $ -1.3 color cut should result a clean sample of stars. However, a detailed selection of specific types of stars needs further refinements.

\section*{Conclusions}

We focused on creating large area galaxy maps of low stellar contamination and high galaxy completeness based on the joint analysis of WISE and 2MASS photometric datasets. Using 2MASS colors add useful information, while $\sim 93 \%$ of the WISE objects with $W1<15.2$ mag have 2MASS pairs. We performed star-galaxy separation using a class of wide-spread machine learning tools, Support Vector Machines. WISE-2MASS objects were cross-identified with SDSS objects, and available SDSS PhotoObj classification data were used as training and control sets. Exhaustive testing of the SVM algorithm with different parameters and inputs revealed that a simple W1-J photometric color cut produces similarly clean data set as the SVM classification, at the expense of loosing $<15 \%$ of the galaxies (Table 1). 

We produced a clean galaxy sample with $1.8\%$ stellar contamination reaching $78.6\%$ completeness using our basic W1$\leq$15.2 and W1-J$\leq$-1.7 cuts. The SVM techniques for the same data yield $3.1\%$ stellar contamination, and $93.4\%$ galaxy completeness. We found at all magnitude limits that SVM algorithms result in slightly higher stellar contamination (an extra $\sim1.5~\%$), but with notable gain in the total number of galaxies identified properly ($\sim10~\%$).

Contamination and completeness are, however, not the only relevant properties of a high-quality galaxy catalog. We probed the isotropy and homogeneity of the resulting galaxy maps, and empirically found that the faintest 2MASS objects at $J>16.5$ mag show inhomogeneities in their density on the sky, presumably due to the survey strategy. We thus supplemented our standard flux and color selection criteria by a $J<16.5$ mag cut in order to produce a more uniform catalog. The resulting catalog contains $N_{\rm gal} \approx 2.4$ million objects, with an estimated star contamination of $1.2\%$, and  $70.1\%$ galaxy completeness. SVM estimates show $2.8\%$ contamination, and $85.9\%$ completeness for this shallower data. These examples demonstrate that the computationally expensive SVM approach creates a more complete catalog with higher stellar contamination.

Regardless the methods applied, the resulting galaxy catalogs represent significant improvement over previous samples using WISE colors only for selection \citep{KovacsEtAl2013}. We not only trace a much larger region in the WISE color-color space than previous catalogs, but also restrict the galaxy selection for only the least noisy W1 band for WISE.

The need for high completeness while minimizing contamination is clear, although we argue that there is no optimal galaxy map in general, since specific science drivers will require different balance between these two effects and thus the cuts that control them. 
Nevertheless, when clean catalog is needed with a simple definition \citep{KovacsEtAl2013,SzapudiEtal2014,FinelliEtal2014}, we recommend W1-J$\leq$-1.7 color cut as a compromise between simplicity, completeness, and the lowest possible stellar contamination. Other applications, e.g., galaxy cluster counting,  gravitational wave source follow-up, etc. might benefit even from denser galaxy maps with higher contamination, and/or non-uniform coverage.

In the near future, we will add photometric redshifts to our catalog matching with SuperCOSMOS \citep{supercosmos}, extending \cite{bilicki2014}. While we plan to make this value added WISE-2MASS-SuperCOSMOS catalog public, at present the WISE-2MASS catalogs can be easily downloaded from the IRSA web-site\footnote{\texttt{http://irsa.ipac.caltech.edu/}} using the queries quoted earlier in this paper.

\section*{Acknowledgments}
AK takes immense pleasure in thanking the support of OTKA through grant no. 101666. In addition, AK acknowledges support from Campus Hungary fellowship program. IS acknowledges support from NASA grants NNX12AF83G and NNX10AD53G. We thank Istv\'an Csabai for useful comments improving the SDSS-WISE-2MASS matching properties, and target selection. We thank the constructive comments by the reviewer of our paper, Katarzyna Malek. Funding for this project was partially provided by the Spanish Ministerio de Econom'a y Competitividad (MINECO) under project Centro de Excelencia Severo Ochoa SEV-2012-0234.
We use HEALPix \citep{healpix}. This publication makes use of data products from the Wide-field Infrared Survey Explorer, which is a joint project of the University of California, Los Angeles, and the Jet Propulsion Laboratory/California Institute of Technology, funded by the National Aeronautics and Space Administration, and data products from the Two Micron All Sky Survey, which is a joint project of the University of Massachusetts and the Infrared Processing and Analysis Center/California Institute of Technology, funded by the National Aeronautics and Space Administration and the National Science Foundation. This work also makes use of data products of the GAMA survey\footnote{\texttt{http://www.gama-survey.org/}}, and the Sloan Digital Sky Survey\footnote{\texttt{http://www.sdss.org/}}.

\bibliographystyle{mn2e}
\bibliography{refs}
\end{document}